# PRECISAMOS DE UMA CONTABILIDADE AMBIENTAL PARA AS "AMAZÔNIAS" PARAENSE?

Ailton Castro Pinheiro[1]

## INTRODUÇÃO

Segundo Leff (2011), o mundo iniciou um processo de consciência ambiental ainda na década de 60 com a *Primavera Silenciosa*, da bióloga Rachel Carson. Em linhas gerais, o livro de Carson explicou como o uso desenfreado de pesticidas nos EUA alterava os processos celulares das plantas, reduzindo as populações de pequenos animais e colocando em risco a saúde humana (BONZI, 2013).

Essa consciência se intensificou mais na década de 70, quando o esgotamento de determinados recursos naturais ficaram mais evidentes (DIAS, 2009). Na década de 70 e posteriores, outros contextos são importantes a serem citados, como a Guerra Fria, a Globalização, a Crise do Petróleo (MITSCHEIN e CHAVES, 2013).

A questão ambiental logo possibilitou uma primeira Conferência entre as Nações em Estocolmo, capital da Suécia. Várias outras Conferências foram realizadas e no debate político e acadêmico surgiram o termo Eco-desenvolvimento, atribuído a Ignacy Sachs, que depois foi substituído pelo termo Desenvolvimento Sustentável (LIMA, 2003).

Estes contextos, contribuíram para a emergência e fortalecimento de múltiplas correntes de pensamentos sobre desenvolvimento sustentável, como o desenvolvimento humano, o desenvolvimento enquanto felicidade, o desenvolvimento regional, o desenvolvimento local, a economia solidária, o desenvolvimento territorial, dentre outras (LEFF,2011; ROCHA, et al, 2016; MITSCHEIN, et al, 2013; DALLABRIDA et al,2004; BENKO e PECQUEUR,2001; DIAS,2009). Isso mostra, que no século XXI vivemos um pluralismo de abordagens teóricas sobre o desenvolvimento, que adentra também nos cenários políticos e influencia a Contabilidade.

A Contabilidade aparece nesse cenário a partir da agenda 21, assinada no âmbito da Conferência das Nações Unidas sobre o Meio Ambiente e Desenvolvimento, realizada no Rio de Janeiro em 1992. Segundo Ferreira, isso ocorre pela necessidade de que países e organismos internacionais desenvolvessem um sistema de contabilidade que integrassem as questões sociais, ambientais e econômicas. A nossa curiosidade nos levou a buscar a compreensão de como a contabilidade "fortemente influenciada pela corrente neoclássica de desenvolvimento" (MAJOR,2017), estruturou seus conceitos para contribuir com o desenvolvimento sustentável? E por vivermos na Amazônia, um espaço tão diversificado, traçamos mais um questionamento: em que medida, o que construímos na contabilidade ambiental brasileira pode ser utilizado para a construção de um desenvolvimento sustentável na Amazônia?

---

[1] Doutorando pelo Núcleo de Meio ambiente da UFPA(NUMA/UFPA).

Como objetivos queremos: **compreender os conceitos de Contabilidade Ambiental no Brasil; fazer críticas e proposições ancorados na realidade ou demanda de uma contabilidade ambiental para Amazônia Paraense**. A estratégia metodológica foi uma análise dos livros de Ferreira (2007); Ribeiro (2010) e Tinoco e Kraemer (2011), onde posteriormente, realizamos correlações destes autores com algumas produções científicas de pesquisadores da Amazônia paraense, especialmente os do Núcleo do Meio Ambiente (NUMA/UFPA) e do Núcleo de Altos Estudos Amazônicos (NAEA/UFPA). Além disso, nossa experiência enquanto pesquisadores deste território, nos ajudou na construção de críticas e proposições para pensarmos uma contabilidade ambiental para as "Amazônias" (TOURNEAU e CANTO,2019).

## RESUMO DOS RESULTADOS E DISCUSSÕES

### Compreensão dos conceitos de Contabilidade Ambiental

A leitura das obras de Ferreira (2007); Ribeiro (2010); Tinoco e Kraemer (2011), possibilitaram a nossa compreensão das seguintes questões temáticas indicadas no quadro a seguir.

| Nº | PERGUNTA TEMÁTICA |
|---|---|
| PNº 01 | Como surgiu a Contabilidade Ambiental? |
| PNº 02 | O que significa Contabilidade Ambiental? |
| PNº 03 | Como a Contabilidade mensura a interação das organizações com a natureza e quais as suas limitações nesse processo? |
| PNº 04 | Como se classifica a Contabilidade Ambiental? |
| PNº 05 | As empresas poluidoras divulgam informações contábeis de caráter ambiental? |
| PNº 06 | O que são Ativos Ambientais? |
| PNº 07 | O que são Passivos Ambientais? |
| PNº 08 | O que são Despesas Ambientais? |
| PNº 09 | O que são Receitas Ambientais? |
| PNº 10 | O que são Custos Ambientais e qual a sua relação com o preço? |

Fonte: elaborado pelo autor.

**Críticas ao modelo neoclássico da Contabilidade Ambiental**

Observando todos estes constructos, não há dúvida de que todos são apenas uma adaptação da contabilidade de princípios neoclássicos e com forte influência da "adoção obrigatória" das Normas Internacionais de Contabilidade no Brasil e no mundo (GORDON,2019).

É a "velha" estratégia de mudarmos os nomes, mas não a essência dos conteúdos neoclássicos, foi assim que construímos a contabilidade tributária, a rural, ambiental, terceiro setor, cooperativa, contabilidade de cooperativa de créditos, contabilidade bancária, contabilidade aplicada ao setor público, contabilidade fiscal, contabilidade orçamentária, dentre tantas outras, porém nenhum desses se transformaram em escolas de pensamentos, com capacidade de quebrar paradigmas, reconstruir-se, gerar debates, contraposições. Talvez o debate, o desconforto às críticas seja o caminho que pode nos levar a construirmos novos rumos para uma Contabilidade Ambiental nas Amazônias.

Segundo Major (2017P.173), "Essa dependência da Contabilidade em relação à teoria econômica neoclássica, levou alguns pesquisadores a afirmar que a Contabilidade se tornou uma 'subárea' dessa corrente econômica, confinando-se aos seus objetivos e pressupostos."

O paradigma neoclássico aliado ao positivismo na contabilidade vem sendo duramente criticado atualmente, por diversos pesquisadores da área contábil de diferentes partes do mundo, principalmente da Europa, com os do Reino Unido e Escandinávia, em particular, por pesquisadores que se dedicam a contabilidade gerencial (uma contabilidade não normatizada). Conforme ela, estes pesquisadores alegam que a corrente neoclássica aliada ao positivismo vem causando uma ossificação e à esterilidade da pesquisa em contabilidade (MAJOR,2017).

Um dos pesquisadores mais crítico do paradigma positivista afirma que:

> Para que o conhecimento seja uma fonte de iluminação e não de dogma, ele deve ter uma dinâmica de mudança. Visto de tal perspectiva, a contabilidade, **como prática**, pode ser e deve ser constantemente examinada, reexaminada, interrogada e criticada no mundo do conhecimento. Em vez de ser uma disciplina por conta própria, a contabilidade precisa recorrer a uma variedade de fontes de iluminação e compreensão. Tem sido e deve continuar a ser um local de investigação **interdisciplinar**. (WOPWOOD,2007 *apud* MAJOR, 2017 p. 175 grifos nosso).

Diante de tudo isso, só podemos afirmar que a contabilidade e sua relação com questões ambientais avançaram se essa afirmação tiver relação com o paradigma neoclássico de desenvolvimento sustentável, para as demais correntes de desenvolvimento, ainda nem começamos a construir novos conceitos; talvez "o mito da caverna de Platão" nos motive a reagir a respeito do que nos alertou Major (2017). Talvez essa dependência neoclássica que estagnou a contabilidade ambiental seja fruto da incompreensão sobre o que significa teoria, autores como Niyama e Silva vem afirmando: "a teoria contábil hoje não é um assunto filosófico longe da realidade, em que a discussão estaria focada em terminologias e **correntes de pensamento**, vivemos a teoria muito próxima da prática. " Se pensarmos assim, seremos uma folha ao vento no bojo de um turbilhão das correntes de pensamentos econômicas que se renovam a todo tempo, ou usando outras palavras, ficaremos para sempre presos na "caverna neoclássica. "

**Proposições para pensarmos uma Contabilidade Ambiental na Amazônia Paraense**

Na Amazônia paraense, precisamos começar a inserir na contabilidade, discussões de outras matrizes teóricas sobre o desenvolvimento sustentável. Por exemplo, no NUMA/UFPA e no NAEA/UFPA, programas da Amazônia com larga experiência, indicam algumas abordagens teóricas viáveis para a elaboração de estratégias de desenvolvimento para Amazônia: o Desenvolvimento Local, o Desenvolvimento Territorial, Governança Territorial, Gestão Social, Economia Solidária etc.

Para esta reconstrução, precisaremos abandonar o modelo positivista de fazer pesquisa, porque como afirma Dias Filho e Machado (2012), este tipo de pesquisa na contabilidade serve para "explicar e predizer" e prefere adotar a estatística e a matemática para validar seus resultados. Na Amazônia, estamos caminhando (seja doutorado profissionais ou acadêmicos) para proposições a partir do entendimento, de que não estamos lidando com uma Amazônia, mas com "Amazônias", como percebe-se no livro: *Amazônias Brasileiras: situações locais e evoluções, recém lançado pelo NUMA/UFPA,* neste caso, compreender contextos e a relação pesquisador e objeto, é fundamental, o que não significa que não se possa utilizar a estatística e a matemática de forma complementar.

Precisamos também, abandonar a ideia de uma contabilidade que só pode registrar o que tem valor monetário. Por que a contabilidade não pode discutir e sistematizar Ativos de forma qualitativa? Nos parece que essa justificativa só pode ser defendida em uma contabilidade de corrente neoclássica. No livro: *história geral da contabilidade no Brasil publicado pelo Conselho Federal de Contabilidade em 2008,* Antônio Lopes de Sá mostra que a base do conhecimento contábil, surgiu antes mesmo que o homem tivesse a escrita e pudesse calcular.

Precisamos pensar uma contabilidade que mostre como pequenas empresas, associações, cooperativas, contribuem de forma positiva para o desenvolvimento de uma cidade, de uma comunidade, de uma floresta, do Brasil.

Necessitamos de uma contabilidade gerencial territorial para o Meio Ambiente, que considere os moradores de comunidades e cidades em riscos, como usuários internos e não externos das informações contábeis.

Precisamos inserir novos elementos contábeis nas correntes de desenvolvimento sustentável, que se discutam na Amazônia, ou seja, precisaremos aprender a teorizar onde não se tem teoria construída e consolidada, para isso, é primordial levarmos essa discussão epistemológica para a graduação, por exemplo, no curso de Economia da UFPA, observamos no PPC do curso que utilizam as disciplinas: Economia Amazônica, História do Pensamento Econômico e Evolução de Ideias Sociais, o que em nossa compreensão, uma adaptação disso para a contabilidade na Amazônia seria fundamental, pois desde logo os alunos poderiam entender que a contabilidade é uma ciência fortemente influenciada por contextos econômicos e sociais e dariam a eles capacidade crítica de contribuir, construir e reconstruir arcabouços conceituais. Não é um trabalho de curto prazo, mas precisamos começar.

Estes ajustes são necessários, porque na Amazônia, precisamos, diante do que está evidenciado na contabilidade ambiental neoclássica, precisaremos também teorizar, como diz Loureiro (2018), a teoria é um sistema explicativo para explicar a realidade que foi criada a partir de um contexto específico.

Podemos nos inspirar na história de Pasteur, contada por Bruno Latour no Livro: *a esperança de pandora*. A descoberta de Pasteur só foi possível segundo ele, porque o pesquisador adotou um ponto de vista totalmente diferente das referências teóricas da química de Liebig. Ele só resolveu adotar um ponto de vista diferente, porque teve o primeiro contato com o objeto, ou seja, o estado concreto que Bachelard (1996) se refere. "Graças ao fermento de Pasteur, a bioquímica surgiu. Antes disso, tinha-se que escolher entre a química ou a biologia" (LATOUR, 2017 p.171).

**CONSIDERAÇOES FINAIS**

Nosso objetivo com este trabalho, foi compreender os conceitos de Contabilidade Ambiental no Brasil; fazer críticas e proposições ancorados na realidade ou demanda de uma contabilidade ambiental para Amazônia Paraense. Em linhas gerais, fizemos uma crítica ao modelo neoclássico dos conceitos construídos na contabilidade ambiental e ao positivismo que lhe acompanha. As proposições são no sentido de indicarmos diretrizes para construirmos uma nova pauta técnica e científica para que a contabilidade possa contribuir de fato, com o desenvolvimento sustentável que estamos discutindo na Amazônia a mais de meio século.

**REFERÊNCIAS**